# Concurrence of anomalous Hall effect and charge density wave in a superconducting topological kagome metal

F. H. Yu[1], T. Wu[1], Z. Y. Wang[1], B. Lei[1], W. Z. Zhuo[1], J. J. Ying[1*], and X. H. Chen[1,2,3†]

[1]Hefei National Laboratory for Physical Sciences at Microscale and Department of Physics, and CAS Key Laboratory of Strongly-coupled Quantum Matter Physics, University of Science and Technology of China, Hefei, Anhui 230026, China

[2]CAS Center for Excellence in Quantum Information and Quantum Physics, Hefei, Anhui 230026, China

[3]Collaborative Innovation Center of Advanced Microstructures, Nanjing 210093, People's Republic of China

*E-mail: yingjj@ustc.edu.cn
†E-mail: chenxh@ustc.edu.cn

**As one of the most fundamental physical phenomena, the anomalous Hall effect (AHE) typically occurs in ferromagnetic materials but is not expected in the conventional superconductors. Here, we have observed a giant AHE in kagome superconductor $CsV_3Sb_5$ with transition temperature (Tc) of 2.7 K. The anomalous Hall conductivity reaches up to $2.1 \times 10^4$ $\Omega^{-1}$ cm$^{-1}$ which is larger than those observed in most of the ferromagnetic metals. Strikingly, the emergence of AHE exactly follows the higher-temperature charge-density-wave (CDW) transition with $T_{CDW}$ ~ 94 K, indicating a strong correlation between the CDW state and AHE. Furthermore, AHE disappears when the CDW transition is completely suppressed at high pressure. The origin for AHE is attributed to enhanced skew scattering in CDW state and large Berry curvature arose from the kagome lattice. These discoveries make $CsV_3Sb_5$ as an ideal platform to study the interplay among nontrivial band topology, CDW and unconventional superconductivity.**

A kagome lattice, representing a two-dimensional network with corner-sharing triangles, provides a fertile ground to study the frustrated, novel correlated and topological electronic states owing to its unusual lattice geometry[1-3]. In general, kagome lattice naturally possesses Dirac dispersion and flat bands that promote electronic correlation effect [4]. Thus, in principle, kagome lattice can exhibit a large variety of electronic instabilities. Many exotic quantum phenomena have been observed in magnetic kagome metals, including giant anomalous Hall effect (AHE) [5-12], chiral edge state[13,14], and topological surface Fermi arcs[15,16]. Up to now, exploring exotic properties in kagome lattice remains quite challenging, particularly for multiple electronic orders.

Recently, a new family of quasi-two-dimensional kagome metals $AV_3Sb_5$ (A = K, Rb, Cs) have attracted tremendous attentions[17]. These materials crystallize in the *P*6/mmm space group with ideal kagome nets of V atoms which are coordinated by Sb atoms. The kagome layers are sandwiched by extra antimonene layers (Sb2) and Cs layers as shown in Figure 1(a).The resistivity of $AV_3Sb_5$ family exhibits anomalies at $T^*$ ranged from 78 K to 110 K, which are ascribed to the formation of charge density wave (CDW) order[17,18]. The observation of superconductivity in the stoichiometric $AV_3Sb_5$ with kagome lattice [19-21]makes this family as an ideal playground to study the correlation among superconductivity, CDW and nontrivial band topology. For example, topological charge order was recently reported in $KV_3Sb_5$ [18] and signatures of spin-triplet superconductivity were also claimed in $Nb-K_{1-x}V_3Sb_5$ devices[22]. Recent reports have shown a unconventional giant AHE in non-superconducting $K_{1-x}V_3Sb_5$ which is attributed to the skew scattering of spin cluster[23]. However, no signature of long-ranged magnetic order or localized magnetic moments has been observed in a near-stoichiometric sample by muon spin resonance (μSR) measurements [24]. One explanation for the aforementioned discrepancy of AHE is attributed to the vacancies at *A* sites[20]. These vacancies might lead to additional local moments around vacancies that gives rise to the giant AHE but are detrimental for the superconductivity. In fact, superconductivity in this family seems to be very fragile by involving the vacancies of alkali metal[20]. Therefore, whether such large AHE is universal for $AV_3Sb_5$ family and its correlation with the emergent orders needs to be elucidated. Here, we observed the giant AHE in the superconducting $CsV_3Sb_5$ with the highest $T_C$ of 2.7 K in the $AV_3Sb_5$ family. The anomalous Hall conductivity reaches up to $2.1 \times 10^4$ $\Omega^{-1}$ $cm^{-1}$, which is larger than those observed in most of the ferromagnetic metals[25]. More interestingly, it is clearly evidenced that the AHE develops spontaneously with the occurrence of CDW, indicating a strong correlation between AHE and CDW state. Such concurrence of CDW and SC behavior is also confirmed in the high-pressure experiments.

Single crystals of $CsV_3Sb_5$ were synthesized via a self-flux growth method similar to the previous reports[19]. In order to prevent the reaction of Cs with air and water, all the preparation processes were performed in an argon glovebox. After high-temperature reaction in the furnace, the excess flux is removed by using water and millimeter-sized single crystal can be obtained. The as-grown $CsV_3Sb_5$ single crystals are stable in the air. X-ray diffraction data are collected at room temperature by using an X-ray diffractometer (SmartLab-9, Rikagu) with Cu K$_\alpha$ radiation. Electrical transport and heat capacity measurements were carried out in a Quantum Design physical property measurement system (PPMS-14T). The longitudinal resistance and Hall resistance were simultaneously measured with standard six-probe Hall bar geometry. Magnetization measurements were performed in SQUID magnetometer (MPMS-5 T, Quantum Design). Piston cylinder cell (PCC) was used to generate hydrostatic pressure above 2 GPa for the high-pressure electrical transport measurements. Daphne 7373 was used as the pressure transmitting medium in PCC. The pressure values in PCC were determined from the superconducting transition of Sn[26].

The crystal structure of $CsV_3Sb_5$ is presented in Fig. 1(a). Fig. 1(b) displays the X-ray diffraction pattern of a $CsV_3Sb_5$ single crystal. Only (00l) diffraction peaks can be detected, indicating the pure phase of the as-grown single crystal with a [001] preferred orientation. The *c*-axis lattice parameter is determined to be 9.308 Å, which is consistent with the previously reported value of $CsV_3Sb_5$[17]. As shown in Fig. 1 (c)-(e), the resistivity, magnetic susceptibility and heat capacity all show a phase transition around 94 K which is ascribed to a CDW-like transition. Meanwhile, bulk superconductivity with $T_C$ of 2.7 K is also observed, being consistent with previous measurements[19]. The residual-resistivity-ratio is 69, indicating the high quality of the measured single crystal. The large anisotropy of the upper critical field (See Fig. S1 in the supplementary material[27]) indicates a 2D nature, which is consistent with the predicted 2D electronic structure in $CsV_3Sb_5$ [19].

To further elucidate its electronic properties, we have performed magnetoresistance measurement under external magnetic field along both the *c*-axis and the *ab*-plane, and the results are shown in Fig. 2(a) and (b), respectively. Large MR(=(ρ(H)-ρ(0))/ρ(0)) and clear Shubnikov-de Haas (SdH) quantum oscillations (QOs) can be clearly observed at low-temperatures with external magnetic field along the *c*-axis. In principle, the Fourier transform of QOs can provides useful insight into the Fermi surfaces. As shown in the inset of Fig. 2(c), the MR curves above 2 Tesla can be well fitted by using a polynomial formula. After subtracting a slowly changed background, the oscillating parts of resistivity Δρ$_{xx}$ as a function of 1/(μ$_0$H) at several representative temperatures are extracted and shown in Fig. 2(c). The fast Fourier transform (FFT) of the QOs reveals four principal frequencies at 18, 26, 72 and 92 T. In contrast, only two frequencies were detected in previous reports on K$_{1-x}$V$_3$Sb$_5$ and RbV$_3$Sb$_5$ samples[21,23]. Since F1 and F2, F3 and F4 are rather close with each other, the doubling of the frequency numbers can be attributed to the slightly warping of FSs along *c*-axis direction. All these frequencies are smaller than that observed in the K$_{1-x}$V$_3$Sb$_5$ and RbV$_3$Sb$_5$ samples, indicating that CsV$_3$Sb$_5$ has the smallest extremal orbits of Fermi surfaces in this family. The effective mass can be extracted from the temperature dependence of the amplitude of FFT peak using the Lifshitz-Kosevich (LK) formula[28]. The oscillation amplitude at a fixed magnetic field is proportional to the thermal damping factor R$_T$: $R_T = \frac{\alpha m^* T}{B \sinh(\alpha m^* T/B)}$, where α = 14.69 T/K is a constant, B = μ$_0$H is the magnetic flux density (taking the average value of the field window used for the FFT of QOs), and m* = m/m$_e$ is the cyclotron mass ratio (m$_e$ is the mass of free electron). The temperature dependence of FFT amplitude of F1 and F2 can be fitted very well as shown in the inset of Fig.2(d), and yields effective mass to be 0.028m$_e$ and 0.031m$_e$ for F1 and F2 orbitals, respectively. The small extremal cross sections of FSs accompanying with light effective mass could be related to the highly dispersive bands in CsV$_3$Sb$_5$.

In order to check the AHE in CsV$_3$Sb$_5$, we have performed the Hall effect measurements at various temperatures as shown in Fig. 3(a) and 3(b). ρ$_{xy}$ exhibits a linear behavior at high temperature, indicating that the carrier density is dominated by one band. ρ$_{xy}$ gradually deviate the linear behavior and the slope of high-field ρ$_{xy}$ gradually decreases below T* and changes its sign around 35 K. The magnitude of Hall coefficient reaches its maximum value around T*, as shown in the inset of Fig. 3(b), similar to the cases of K$_{1-x}$V$_3$Sb$_5$[23] and RbV$_3$Sb$_5$[21]. Since CsV$_3$Sb$_5$ is a multiband system, such behaviors could be attributed to the significant enhancement of hole mobility at low temperature[21]. In the low-field region, highlighted by the gray shading area in Fig. 3(a), an antisymmetric sideways "S" line-shape is observed, which is similar to that reported in the K$_{1-x}$V$_3$Sb$_5$ sample. We extract $\rho_{xy}^{\mathrm{AHE}}$ by subtracting the local linear ordinary Hall term (typically from 1T to 2T at low temperature) as shown in the Fig. 3(c)-(d). $\rho_{xy}^{\mathrm{AHE}}$ decreases with increasing the temperature and persists up to the CDW transition temperature T*. We also obtain the Hall conductivity by inverting the resistivity matrix, σ$_{xy}$ = -ρ$_{xy}$/(ρ$_{xx}^2$ + ρ$_{xy}^2$). The anomalous Hall conductivity (AHC) σ$_{AHE}$ is obtained by subtracting linear local ordinary Hall conductivity background (details see Fig. S4 in the supplementary material[27]). As shown in Fig. S2 of the supplementary material[27], the angle-dependent σ$_{AHE}$ can't be linearly scaled with an out-of-plane component of magnetic field, which excludes normal Hall effect as a possible origin.

One remarkable observation here is that the AHE emerges simultaneously with the CDW transition. To confirm such correlation between CDW and AHE is intrinsic rather than coincident in this type of materials. We use high pressure to directly tune the T* and check whether AHE is concurrence with CDW order. In Fig. 4a, T* can be suppressed to 33 K with the pressure around 1.42 GPa, similar to the previous high-pressure results[29,30]. With temperature below T*, the hall resistivity exhibits an antisymmetric sideways "S" line-shape in the low-field region as shown in Fig. 4b, which is the same with the case at ambient pressure. The anomalous Hall conductivity has been extracted as shown in Fig. 4c, which clearly shows the sudden appearance of the AHE below T*. Further increasing the pressure to 2.02 GPa, the CDW can be completely suppressed as shown in Fig. S5 of the supplementary material[27]. At the meantime, the AHE disappears and the Hall resistivity can be well fitted by using the two-band model as shown in Fig. S7. Our high-pressure transport measurements confirm that the CDW and AHE occur at the same time similar to the case at the ambient pressure.

With temperature decreasing below T*, $\rho_{xy}^{AHE}$ at ambient pressure exhibits a rapid increasing around 20~40 K as shown in Fig. 5(a). Such a behavior is qualitatively similar to that previously reported in $K_{1-x}V_3Sb_5$. The MR at 14 T is plotted as a function of temperature as shown in Fig. 5(b). The MR shows a sudden jump below T* indicating its strong correlation with the CDW order. The carrier mobility is enhanced and the scattering can be effectively suppressed when entering CDW state, thus giving rise to the enhancement of MR. More interestingly, below T*, $\sigma_{AHE}$ can be well scaled with MR as shown in Fig. 5(b). All these discoveries indicate that the origin of AHE should correlate strongly with CDW order in the $AV_3Sb_5$ system. To further understand the AHE in $CsV_3Sb_5$, we have plotted $\sigma_{AHE}$ versus $\sigma_{xx}$ together with a variety of other typical AHE materials as shown in Fig. 5(c)[5,6,8,10,23,31,32]. Various AHE regimes from the localized hopping regime to the skew scattering regime are divided by considering the value of $\sigma_{xx}$. $\sigma_{AHE}$ follows quadratic scaling with $\sigma_{xx}$ below 45 K and locates at almost the same position with $K_{1-x}V_3Sb_5$. However, at higher temperature, $\sigma_{AHE}$ severely deviates from the $\sigma_{xx}^2$ behavior. Meanwhile, the $\sigma_{AHE}$ for $CsV_3Sb_5$ at 1.42 GPa also locates at the same position with that at ambient pressure. These results indicate that AHE is universal in $AV_3Sb_5$ family. It may be from complex origins by involving both Berry curvature and skew scattering. Next, we will discuss this issue with more details.

In a short summary, a giant $\sigma_{AHE}$ is revealed in a kagome superconductor $CsV_3Sb_5$, which is the first case of AHE material which shows bulk superconductivity to the best of our knowledge. Usually AHE occurs with broken time-reversal symmetry, typically in a ferromagnetic phase which is not favored for conventional superconductivity. The discovery of AHE in the superconducting kagome metal indicates either superconductivity or AHE hosts an unconventional mechanism in this type of materials. AHE is usually ascribed to the Berry curvature of band structure in magnetic phase or electron scattering due to magnetic impurities [25,33]. According to the underlying mechanism, the AHE can be roughly divided into intrinsic and extrinsic categories. The intrinsic AHE is related to the band structure and proportional to the integral of Berry curvatures below Fermi level. Large intrinsic AHE has been widely observed in the magnetic topological materials such as $Co_3Sn_2S_2$[8,9] and $Co_2MnGa$[11,12]. In our case, when CDW transition occurs, a gap may open on the topological band which could introduce a large Berry curvature in this system, leading to the giant AHE. In fact, scanning tunneling microscopy measurements find evidence that the CDW state displays a chiral anisotropy with an energy gap opening at Fermi level[18]. However, the calculated intrinsic

AHE based on realistic parameters is much smaller than the AHE observed in our experiment[18], indicating that intrinsic AHE mechanism alone can't give a satisfactory explanation. The extrinsic AHE depends on asymmetric electrons scattering in the periodic potential of a crystal, which is usually caused by magnetic impurities[25]. The skew scattering from disorder tends to dominate the AHE in highly conductive ferromagnets. The giant AHE at low temperature in CsV$_3$Sb$_5$ locates in the skew scattering region and follows the $\sigma_{xx}^2$ behavior as shown in Fig. 5(c), indicating that the main contribution of AHE at low temperature comes from the skew scattering[23,34,35]. Since orbital magnetoresistance also depends on the electron scattering, it can qualitatively explain the well scaling behavior between MR and AHE. Since the superconducting CsV$_3$Sb$_5$ is likely nonmagnetic and local moment is absent in KV$_3$Sb$_5$ indicated by the μSR experiments[24], the mechanism of skew scattering in this system is still not well understood. However, tiny amounts of paramagnetic impurities may exist in the single crystal judging from the Curie-Weiss tail in the magnetic susceptibility curve below 50 K (Fig.1d), almost the same temperature below which σ$_{AHE}$ rapidly increases and follows quadratic scaling with σ$_{xx}$. By fitting the susceptibility at low temperatures with the Curie-Weiss law, we can estimate the level of paramagnetic impurity to be 0.7%. It would be interesting for the theorists to figure out whether a large extrinsic AHE could origin from such low level of paramagnetic impurities in a kagome metal. The unusual lattice geometry of the kagome lattice intertwined with CDW state could be the possible origin of large skew scattering. Quantitatively understanding for the giant AHE observed in *A*V$_3$Sb$_5$ family and its correlation with superconductivity, CDW and topological states still needs further investigations.

In conclusion, we observed a giant AHE in superconducting CsV$_3$Sb$_5$ single crystals with kagome lattice. The concurrence of AHE and CDW indicates the unconventional origin of AHE, which may be possibly relate to the enhanced skew scattering in CDW state and large Berry curvature due to kagome lattice. Our work suggests unconventional superconductivity in CsV$_3$Sb$_5$ and stimulates broad interests in studying the correlation between CDW state and AHE in the kagome materials.

ACKNOWLEDGMENTS

This work was supported by the National Key Research and Development Program of the Ministry of Science and Technology of China (Grants No. 2019YFA0704901 and No. 2017YFA0303001), the Anhui Initiative in Quantum Information Technologies (Grant No. AHY160000), the Science Challenge Project of China (Grant No. TZ2016004), the Key Research Program of Frontier Sciences, CAS, China (Grant No. QYZDYSSWSLH021), the Strategic Priority Research Program of the Chinese Academy of Sciences (Grant No. XDB25000000), the National Natural Science Foundation of China (Grants No. 11888101 and No. 11534010), and the Fundamental Research Funds for the Central Universities (WK3510000011 and WK2030020031)

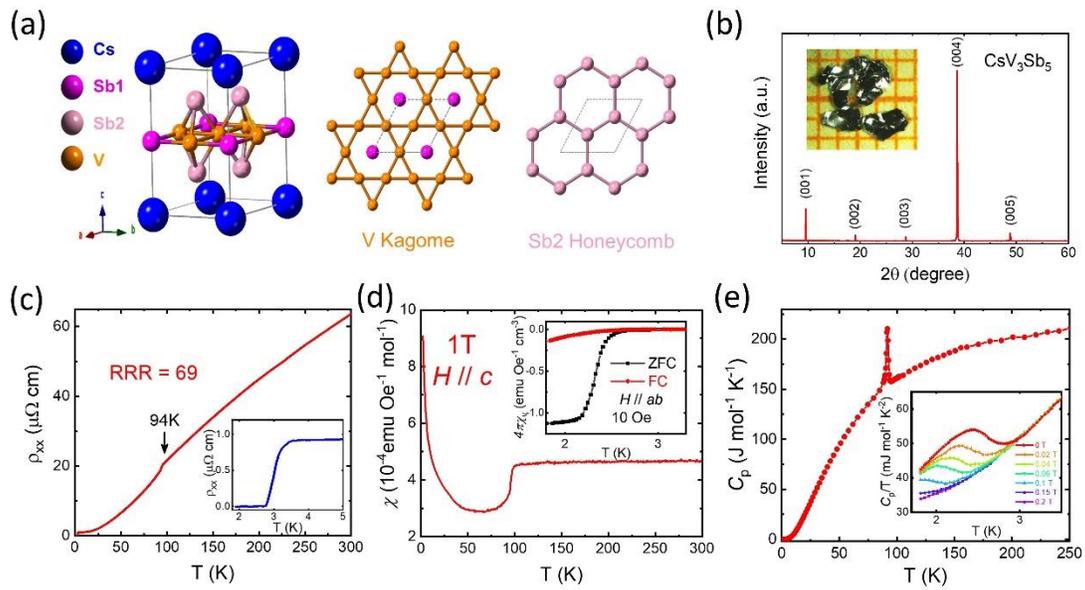

Figure 1. (a): Crystal structure of $CsV_3Sb_5$. The vanadium sublattice forms perfect kagome lattice. There are two distinct Sb sublattices. Sb1 atom is centered on each kagome hexagon and Sb2 sublattice creates an antimonene layer below and above each kagome layer. (b): X-ray diffraction pattern of $CsV_3Sb_5$ single crystal with the corresponding Miller indices (00L) in parentheses. The inset shows photo of typical $CsV_3Sb_5$ single crystals on a 1 mm grid paper. (c), (d) and (e): Temperature dependence of resistivity, magnetization and heat capacity of $CsV_3Sb_5$ single crystal. Clear phase transition around 94 K was observed in all measurements, possibly related to the charge density wave. The insets represent the bulk superconductivity around 2.7 K in all the measurements.

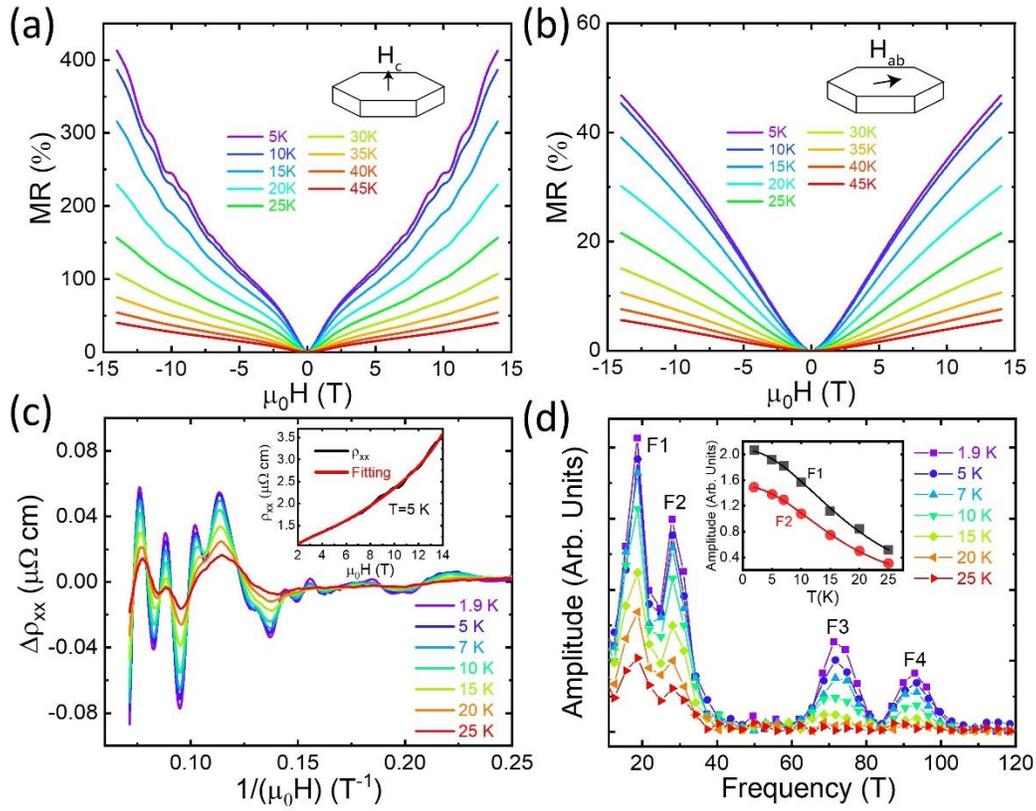

Figure 2. Magnetoresistance measured at various temperatures with magnetic field along *c* axis (a) and in *ab* plane (b), respectively. (c): Magnetoresistance after subtracting a monotonic background plotted as a function of $1/(\mu_0H)$ at various temperatures. The inset shows a polynomial fitting of a MR curve above 2 T. (d) Extracted fast Fourier transform frequency showing four principal frequencies. The inset shows the Lifshitz-Kosevich fit of the F1 and F2 orbits to extract the effective mass.

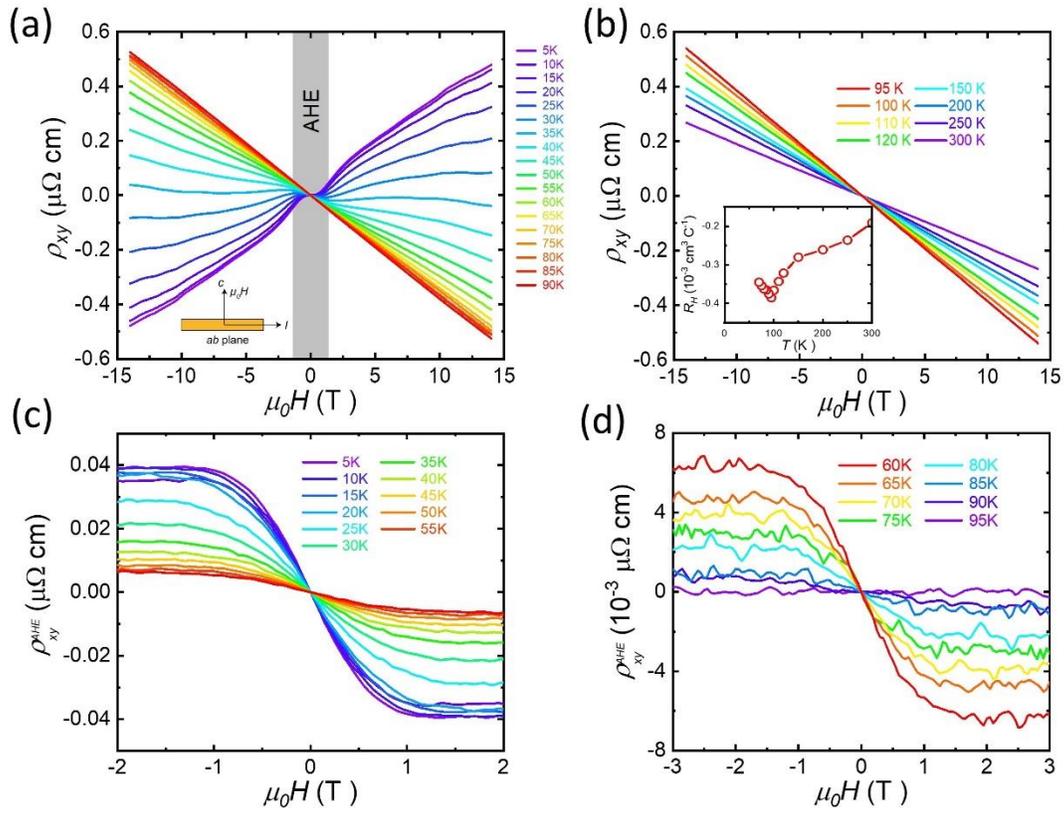

Figure3. (a) and (b): Field dependence of Hall resistivity at various temperatures with magnetic field up to 14 T. The gray area represents the AHE in the low-field region. The inset shows the temperature dependence of Hall coefficient. The magnitude of $R_H$ reaches its maximum value around T*. (c) and (d): Extracted $\rho_{xy}^{AHE}$ by subtracting the local linear ordinary Hall background at various temperatures. AHE spontaneously emerges below T*.

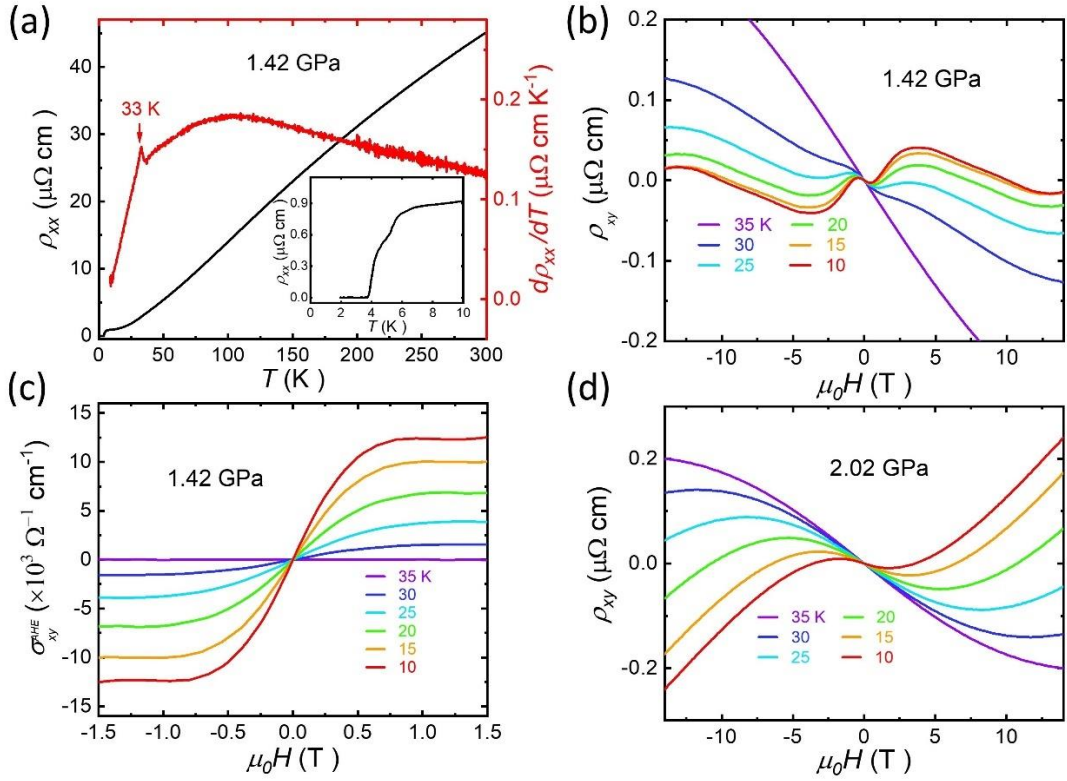

Figure4. (a): Temperature dependence of the resistivity curve for $CsV_3Sb_5$ at 1.42 GPa. T* is suppressed to 33 K as determined from the derivative resistivity curve. The inset exhibits the superconducting transition at 1.42 GPa. (b): Hall resistivity at various temperatures under the pressure of 1.42 GPa. The AHE sudden appears below T*. (c): The extracted anomalous Hall conductivity at various temperatures. (d): The Hall resistivity at various temperatures with the pressure of 2.02 GPa. No AHE is observed since CDW is completely suppressed.

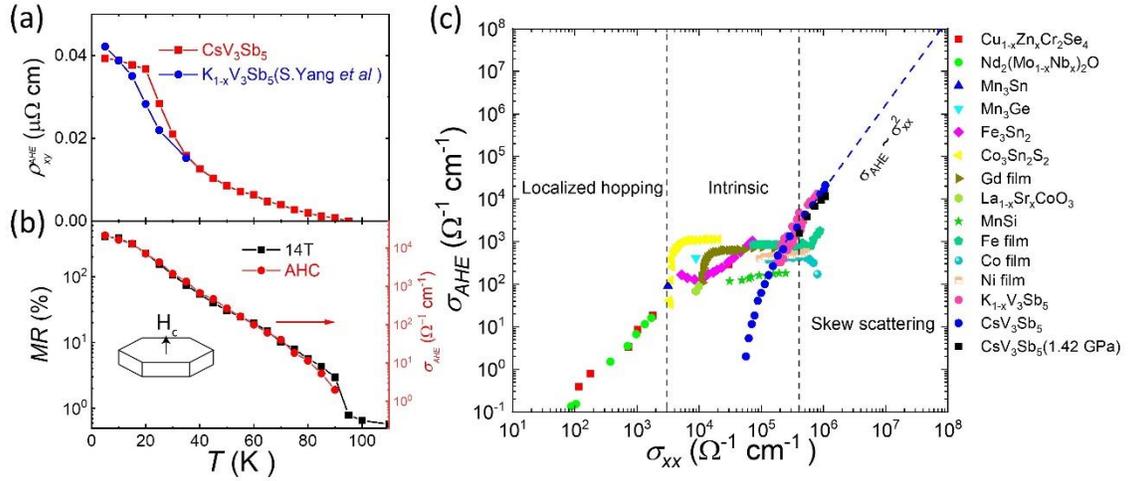

Figure5. (a): The $\rho_{xy}^{AHE}$ as a function of temperature for CsV$_3$Sb$_5$ and K$_{1-x}$V$_3$Sb$_5$. The data of K$_{1-x}$V$_3$Sb$_5$ is taken from Ref.[23]. (b): Scaling behavior of magnetoresistance at 14 T with σ$_{AHE}$ for CsV$_3$Sb$_5$. (c): σ$_{AHE}$ versus σ$_{xx}$ for a variety of materials comparing with CsV$_3$Sb$_5$ spanning various regimes from localized hopping regime to the skew scattering regime. For CsV$_3$Sb$_5$, σ$_{AHE}$ follows quadratic scaling with σ$_{xx}$ only at low temperature and locates at the same position with K$_{1-x}$V$_3$Sb$_5$.